\documentclass[letterpaper]{article}
\usepackage{jheppub}
\pdfoutput=1
\usepackage{color,latexsym, array,multirow,  verbatim}
\usepackage[config, singlelinecheck=true]{caption}
\usepackage{url}
\usepackage{multirow}
\usepackage{subfigure,color}

\definecolor{darkred}{rgb}{0.5,0.0,0.0}

\newcommand{\mbb}{m_{b\bar{b}}}

\def\simgt{\mathrel{\lower2.5pt\vbox{\lineskip=0pt\baselineskip=0pt
           \hbox{$>$}\hbox{$\sim$}}}}
\def\simlt{\mathrel{\lower2.5pt\vbox{\lineskip=0pt\baselineskip=0pt
           \hbox{$<$}\hbox{$\sim$}}}}

\newcommand{\beq}{\begin{equation}}
\newcommand{\eeq}{\end{equation}}
\newcommand{\beqs}{\begin{eqnarray}}
\newcommand{\eeqs}{\end{eqnarray}}

\keywords{}
\preprint{}

\begin{document}

\title{
Quantifying the power of multiple event interpretations}

\author{
Yang-Ting Chien$^a$, David Farhi$^b$, David Krohn$^b$, Andrew Marantan$^b$, David Lopez Mateos$^b$, Matthew Schwartz$^b$}

\affiliation{$^a$Theory Division T-2, Los Alamos National Laboratory, Los Alamos, NM 87545}
\affiliation{$^b$Department of Physics, Harvard University, Cambridge MA, 02138}
\emailAdd{ytchien@lanl.gov}
\emailAdd{farhi@physics.harvard.edu}
\emailAdd{dkrohn@physics.harvard.edu}
\emailAdd{amarantan@physics.harvard.edu}
\emailAdd{david.lopez@cern.ch}
\emailAdd{schwartz@physics.harvard.edu}

\abstract{
A number of methods have been proposed recently which exploit multiple highly-correlated interpretations
of events, or of jets within an event. For example, Qjets reclusters a jet multiple times and telescoping
jets uses  multiple cone sizes. Previous work has employed these methods in  pseudo-experimental analyses and found that, with a simplified statistical treatment, they give sizable improvements over traditional methods. 
In this paper,  the improvement gain from multiple event interpretations is explored with methods much closer to those used in real experiments.
To this end, we derive a generalized extended maximum likelihood procedure. We study the significance improvement in Higgs to $b \bar{b}$ with both this method and the simplified method from previous analysis. With either method, we find that using multiple jet radii can provide substantial benefit
over a single radius. Another concern we address is that multiple event interpretations
might be exploiting similar information to that already present in the standard kinematic variables. 
By examining  correlations between kinematic variables commonly used in LHC analyses and 
invariant masses obtained with multiple jet reconstructions, we find that using multiple radii is still helpful even on top of standard kinematic variables when combined with boosted decision trees. These results suggest that including multiple event interpretations
in a realistic search for Higgs to $b\bar{b}$ would give additional sensitivity over traditional approaches. 
\vspace{5cm}
}

\maketitle

\section{Introduction}
Both the ATLAS and CMS collaborations have recently released their full Run 1 analyses
of the search for the  $H\rightarrow b\bar{b}$  decay mode~\cite{atlas,cms}. With only Run 1 data, neither search was
capable of finding this process at Standard Model rates. With the additional statistics from Run 2 data, $H\to b\bar{b}$ will surely be observed. 
However, a precision measurement providing a meaningful extraction of the bottom and top Yukawa couplings with
implications for Beyond the Standard Model physics will require these 
searches to increase their sensitivity beyond what might be achievable with currently used
experimental techniques. Most of the proposed improvements involve looking in special kinematic regions where
backgrounds are smaller~\cite{Butterworth:2008iy} or computing new, physically-motivated observables~\cite{Gallicchio:2010dq,Gallicchio:2010sw}.

In~\cite{Ellis:2012sn}, a qualitatively new way to construct observables called ``Qjets'' was proposed. Instead of  comparing a single 
observable between signal and background, Qjets proposed to look at the sensitivity of an observable to multiple interpretations 
of that observable generated by small variations in some parameter. In the original
Qjets proposal, these variations were made perturbing around the original jet clustering algorithm: instead of always 
merging the two
closest particles during jet clustering, the Qjets algorithm considers merging more distant pairs. This generates a distribution of highly-correlated observables
for each jet in each event. The width of this distribution for pruned jet mass~\cite{Ellis:2009me}, which was called {\it volatility} in~\cite{Ellis:2012sn},
provides a strong signal-to-background discriminant in boosted Higgs or boosted $W$ boson searches. Volatility has been measured in experiment~\cite{ATLAS-CONF-2013-087,CMS-PAS-JME-13-006} with similar
results and discrimination power to simulation. An application of the Qjets method to event reconstruction was proposed in~\cite{Kahawala:2013sba}. 
In~\cite{Chien:2013kca}, a simpler and faster way of using multiple event interpretations was proposed: simply compute the same observable
using different values of the jet size $R$. Both \cite{Kahawala:2013sba} and~\cite{Chien:2013kca} computed 
the reach of the $H\to b\bar{b}$ search by combining the multiple interpretations, finding as much as a 46\% improvement in significance over using a single interpretation.

The method used to estimate significance improvement in~\cite{Ellis:2012sn,Kahawala:2013sba,Chien:2013kca} is
a natural generalization of cut-and-count for a single observable. 
Normally, an event either passes a set of cuts ($z=1$) or does not ($z=0$). With multiple interpretations, a fraction $z$ of the interpretations pass the cuts. This fraction $z$ is an observable, measurable in data and 
computable with Monte Carlo. In~\cite{Ellis:2012sn,Kahawala:2013sba,Chien:2013kca}, the 1-dimensional distributions of $z$ for signal and background were
used to estimate the probability that a given set of of events could be explained by a fluctuation of the background only. The procedure is reviewed below
in Section~\ref{sec:measures}.
Using this method, multiple event interpretations were shown in \cite{Ellis:2012sn,Kahawala:2013sba,Chien:2013kca} to give significant improvement to search reaches.

One drawback of the method used in~\cite{Ellis:2012sn,Kahawala:2013sba,Chien:2013kca} is that it presupposes a knowledge of the background
cross section. Many LHC analyses try to avoid taking cross sections from theory. Instead, they often use
control regions to establish background normalizations, which are not necessarily precisely
known in the specific regions of phase space exploited by the analysis. These control regions are 
typically defined to have minimal overlap with the signal regions. 
When using multiple event interpretations, however, it is generally not possible to define 
non-overlapping regions, since a single event can potentially cover a large range of values
for the observable of interest. One goal of this paper is to generalize the extended maximum likelihood procedure,
which fits to signal and background cross sections separately, to observables based on multiple event interpretations.

Another question one might have about multiple event interpretations is whether the improvement is due to features of the
events which are already accounted for in the current search strategies. For example, most analyses (including the $H\rightarrow b\bar{b}$ 
searches) use many kinematic variables in the event to maximize significance, often combined with sophisticated multivariate methods
like neural networks and boosted decision trees. It has not been
proven so far that the improvement observed when using multiple event interpretations is 
independent of what can be obtained by exploiting the kinematic features of the event exploited 
by multivariate discriminants. We also address this concern, by showing that multiple event interpretations can indeed improve the significance of a search when combined with kinematic variables.

This paper is organized as follows. Section~\ref{sec:evtSel} describes the simulation and
event selection we used. 
Section~\ref{sec:measures} gives a quick introduction to the statistical methods we discuss and their relative merits. 
In Section~\ref{sec:Like}  we first review the typical extended
maximum likelihood (EML) fit, emphasizing those
aspects which break down when using multiple event interpretations. Then we derive
 the modifications in the EML formalism necessary to account for the statistical
correlations among the different interpretations of the same event. 
Section~\ref{sec:multLike} describes a two-dimensional extension of the
likelihood fit that avoids modifications of the EML fit at the expense of adding complexity, and
compares the performance of this extension to the results of the previous section.  
Section~\ref{sec:mva} compares the performance of multivariate analyses including
kinematic information and multiple event interpretations, to understand whether multiple event
interpretations indirectly make use of kinematic information already exploited in current 
LHC analyses. We summarize the results from all these methods in Table~\ref{tab:results}.

\section{Monte Carlo simulation and event selection}\label{sec:evtSel}
We generate signal and background processes for proton-proton collisions at 
$\sqrt{s}=8$~TeV using MadGraph 5.1~\cite{Alwall:2011uj}, 
interfaced to Pythia 6.4~\cite{Sjostrand:2006za} to simulate the parton shower and non-perturbative
effects such as underlying event and hadronization. $ZH\rightarrow e^+e^-b\bar{b}$ events 
are generated and used for the signal, and $Z(\rightarrow e^+e^-)+b\bar{b}$ events are 
used as background. High statistics are produced for these samples, but in quoting expected
significances, the signal and backgrounds are normalized to 25~fb$^{-1}$. This normalization
is also used for generating toy models used for estimating the uncertainties in the likelihood fits. 
Jets are clustered from stable particles with lifetimes above 10~ps  
(excluding neutrinos) using the anti-$k_t$ algorithm with different $R$ parameters. Unlike in~\cite{Kahawala:2013sba}, 
these interpretations are built using a deterministic method, similar to the telescoping jets approach introduced in~\cite{Chien:2013kca}. 
The event selection is a simplified version of that in ~\cite{atlas} and 
requires $83$~GeV$<m_{ee}<99$~GeV, $p_T(b_{\rm lead})>45$~GeV, 
$p_T(b_{\rm sublead})>25$~GeV and $p_T^{\rm lep}>25$~GeV. A jet is defined as a $b$-jet
if it contains any decay products from the original $b$-quark. Studies are performed 
in two kinematic regions defined by the transverse momentum of the vector boson 
($p_T^Z<120$~GeV and $p_T^Z>120$~GeV). The invariant mass distribution of the
two leading jets in $p_T$ that are labeled as $b$-quarks is used for our singificance estimates.

\section{Significance measures} \label{sec:measures}
In this section, we quickly review the essential differences between a cross-section based significance calculation, like
the ones used in~\cite{Ellis:2012sn,Kahawala:2013sba,Chien:2013kca}, and the extended likelihood method.

The approach used in~\cite{Ellis:2012sn,Kahawala:2013sba,Chien:2013kca} starts with an observable $z$, defined as the fraction of event interpretations
satisfying a given set of cuts. In a normal analysis, $z$ would be either 1 (event passes cuts) or 0 (event fails cuts). The insight of Qjets was
that one can use multiple event interpretations to make $z$ a rational number. One can then compute in Monte Carlo the probabilities $\rho_S(z)$ and $\rho_B(z)$
for finding certain $z$'s (See Fig. 3 of~\cite{Kahawala:2013sba} for some $\rho(z)$ distributions).
 Then the probability that the data cannot be accounted for by a fluctuation of the background (measured in standard deviations of the signal away from the background) is given by
\begin{equation}
\text{significance}
 = 
\frac{N_{\text{data}} - N^B_{\text{expected}}} {\delta N^B_{\text{expected}}}
= \frac{ \int_0^1 d z \dfrac{\rho_S(z)}{\rho_B(z)} \Big[ \rho_\text{data}(z) - \rho_B(z) \Big]}{
\sqrt{\int_0^1 dz \dfrac{\rho_S(z)^2}{\rho_B(z)}}
}
\label{sigoft}
\end{equation}
where $\rho_{\text{data}}$ is the observed probability. See Section 5 of~\cite{Kahawala:2013sba} for a derivation of this formula. In a simulation, we replace data-minus-expected-background in the numerator by signal:
 $\rho_\text{data}(z) - \rho_B(z) \to \rho_S(z)$. For terminological clarity, we call the estimate of significance using this
method the {\bf cross-section  (xs)-based significance}. It is cross-section based since one needs to know how much background there should be in order
to see a fluctuation above this amount. 

The xs-based significance can be applied to any observable, not just this $z$ variable. 
It corresponds to an analysis which provides a weight function $w(z) = \rho_S(z)/\rho_B(z)$, and measures a final observable $N$ which is the sum of the weight of all observed events. In particular, it can be applied to multidimensional data, using $w(z_1,z_2,\cdots)$ if the multidimensional distributions $\rho_S(z_1,z_2,\cdots)$ and 
$\rho_B(z_1,z_2,\cdots)$ are known. We show the relative significance using various choices of $z_i$ in Table~\ref{tab:results}.

Many LHC analyses are instead based on likelihood fits.
In a likelihood fit, the distribution of some observable $z$ in the data is used to fit the signal and background cross sections:
\begin{equation}\label{fitschematic}
\rho_\text{data}(z) = (\sigma_S \pm \delta_S )\rho_S(z) + (\sigma_B \pm \delta_B)\rho_B(z)
\end{equation}
While the xs-based significance is computed using the normalized probability distributions
for signal and background and the expected background cross section, likelihood fits extract both signal and background cross sections directly from data.
A likelihood fit which takes into account the Poisson fluctuations of signal and background is called an {\bf extended maximum likelihood  (EML) fit}; in the rest of this paper all references to ``likelihood" fits include this extension as appropriate. In Section~\ref{sec:Like} 
we show how extended maximum likelihood can be computed from events with multiple interpretations.

In likelihood fits the ultimate observable of interest is $\sigma_S$. For a discovery, we need the probability that $\sigma_S > 0$. The analog of the significance in Eq. \eqref{sigoft} is the expected value of $\sigma_S$ divided by the variance in the measured value of $\sigma_S$ on data sets with no signal. In other words:
\begin{equation}\label{sigfit}
\text{significance}
=
\frac{\left<\sigma_S\right>_{\text{signal+background}}}{\left<\delta_S\right>_{\text{background only}}}
\end{equation}

Note that while Eqs. \eqref{sigoft} and \eqref{sigfit} are both measures of the probability that a signal is there, the two methods have different priors so they cannot be directly compared.
In the xs-based significance, the analysis knows the expected background rate, and counts any difference in the data from the background as observed signal. For an  EML significance, both the signal rate and the background rate are fit. Since the xs-based analysis has access to more information it will often give higher significances. The likelihood fit is procedurally closer to experimental conditions, where backgrounds are never known with perfect accuracy and are estimated from sidebands.

The difference in priors between xs-based and EML can be illustrated by a simple example. Suppose we had only a single bin, and so the entirety of the data is the fact that $N$ events were measured. The xs-based approach would report an observed value $N$, subtract the expected $N_B$ (an input from theory), and declare an observed signal rate of $N_S=N-N_B$, with a corresponding significance $N_S/\sqrt{N_B}$. The likelihood fit, however, would simply fail, for it would be attempting to fit $N=\sigma_S+\sigma_B$ for both $\sigma_S$ and $\sigma_B$, and a single bin \emph{cannot} fit two parameters. Thus the error would be infinite, and the significance would not be meaningful.

\section{Maximum likelihood fits}\label{sec:Like}

Many methods exist to find the cross section fits $\sigma_S$ and $\sigma_B$ in Eq.~\eqref{fitschematic} by maximizing the likelihood of the observed distributions. In this section we explore how to apply such methods to multiple-interpretation data. 
When using multiple interpretations, each event $k=1\cdots K$ gives us not a single number $x_k$, as is the usual case, but a series of numbers $x_k^i$, with $i=1\cdots I$ indexing the interpretations. 
We consider two approaches to constructing a likelihood fit from these observables $x_k^i$, which we call the merged histogram likelihood fit and the multidimensional likelihood
fit.

For the {\bf merged histogram fit}, we combine all of the interpretations into a single one-dimensional distribution, as though we had not $K$ events but $K\times I$ events. Then we can apply the usual fitting technology to this distribution, as we might for, say, an invariant mass distribution in a conventional analysis. This approach will converge upon the correct fit values in the limit of infinite statistics. However, because the distribution contains both highly-correlated contributions (from a single event) and statistically-uncorrelated ones (from different events), the existing technology to estimate the errors in the fit parameters will give vastly incorrect errors. We show how to correctly estimate the errors with this method in Section \ref{sec:oneDimLL}.

In the {\bf multidimensional likelihood fit}, we treat the $x_k^i$ as $I$ observables in $K$ events. This is possible when the multiple interpretations are indexed,
as in telescoping jets where they correspond to different $R$, but not when they are generated by adding randomness to a jet algorithm, as in the original Qjets
proposal. If the $I$ interpretations are distinguishable, we can then do a fit to the $K$ data points in an $I$-dimensional space. This is no different from a regular fit to multi-dimensional data, and so the usual fitting technology can be applied. As in other multidimensional fits, the high-dimensionality of the space will quickly saturate the statistics either of data or simulation. To control for this, we consider three approaches. We first try severely limiting the number of interpretations, down to $I=2$. This allows us to map out the full $I$-dimensional distribution over events and produce a nearly exact maximum likelihood solution. Second, we try limiting the number of bins, but taking $I$ larger. Third, we try using multivariate methods, in particular boosted decision trees, in place of the exact likelihood. The multidimensional approaches are explored in Section \ref{sec:multLike}. Numerical results are summarized in a Table~\ref{tab:results} and discussed in Section~\ref{sec:results}.

\subsection{Merged histogram likelihood fit}\label{sec:oneDimLL}
In this section we derive appropriate formulas for uncertainty estimation when highly correlated data from multiple event interpretations are merged into
a single distribution. We first review general features of how likelihood fits are done, and then discuss how things are modified with multiple event 
interpretations.

In an extended maximum likelihood (EML) fit, a set of unknown parameters $\{a_\alpha\}$ are estimated
by maximizing the likelihood density over a sample of $N$ data values, $x_n$:
\begin{equation}
\mathcal{L}=\left[\prod_n^{N} P(x_n; \{a_\alpha\})\right]e^{-\mathcal{N}(\{a_\alpha\})},
\label{eq:stdLL}
\end{equation}
where $P(x; \{a_\alpha\})$ is the probability density for the data value $x$ when $\mathcal{N}$ data points
are expected. It should be noted that $N$, the actual number of observed events, is generally
not equal to $\mathcal{N}$, due to Poisson fluctuations ($\mathcal{N}$ can take non-integer values). Since the normalization has to be 
constrained to the number of expected events, the following is true:
\begin{equation}
\int P(x;\{a_\alpha\}) dx=\mathcal{N}(\{a_\alpha\}), 
\end{equation}
where $P$ is taken as a continuous probability density function in this equation and
not evaluated at each data point. 
This shows one way that the EML fit differs from the standard maximum 
likelihood fit, for which this normalization is 1. The EML thus allows to fit for the shape of
a distribution as well as for the normalization. The EML
is thus widely used in searches, where the total number of events collected is subject to
Poisson fluctuations. 

The difference between the standard maximum likelihood and the EML lies on the normalization
condition, represented by the $e^{-\mathcal{N}}$ term~\cite{eml}, which guarantees that 
the normalization of $P(x_n;\{a_\alpha\})$ does not increase arbitrarily in the maximization procedure,
and that that normalization obeys Poisson statistics for $\mathcal{N}$ expected events. 
The log-likelihood for this function takes the form
\begin{equation}\label{onedimlikelihood}
\ln{\mathcal{L}}=\sum_{n=1}^N \ln P(x_n;\{a_\alpha\})-\mathcal{N}(\{a_\alpha\}). 
\end{equation}
The fitted parameters are then taken to be the set $\{\hat a_\alpha\}$ which maximize this likelihood:
\begin{equation}
0 = \frac{d\ln\mathcal{L}}{da_\alpha}\left(\{\hat a_\alpha\}\right)
\end{equation}

The errors on these parameters come from incoherent Poisson fluctuations on the number of observed events in each bin, $\delta n(x)$. Thus the total error on each parameter, $\delta a_\alpha$, is given by the quadrature sum of the errors due to each bin fluctuation:
\begin{equation}
\delta a_\alpha^2 = \sum_{\textrm{bin }x} [\delta a_\alpha\left(\delta n(x)\right)]^2
\end{equation}
The error due to fluctuations of a single bin can be computed by expanding the minimization equation:
\begin{equation}
0 = \frac{d\ln\mathcal{L}}{da_\alpha}\left(n(x) +\delta n, \hat a+\delta \hat a\right) = \frac{d\ln\mathcal{L}}{dn(x)da_\alpha}\delta n(x)+\sum_\beta\frac{d\ln\mathcal{L}}{da_\alpha da_\beta}\delta \hat a_\beta
\end{equation}
$d\ln\mathcal{L}/dn(x)$ is just the contribution to $\ln\mathcal{L}$ from a single event in bin $x$, which is given by $\ln P$. From this we can compute the total error on parameters $a_\alpha$:
\begin{equation}
\delta a_\alpha^2 = \sum_{\textrm{bin }x} \left(\delta n(x)\sum_\beta M_{\alpha\beta}^{-1}\frac{d\ln P}{da_\beta}\right)^2
\end{equation}
Where $M_{\alpha\beta}$ is the second derivative matrix of $\ln\mathcal{L}$. Assuming $n(x)$ has Poisson errors $\delta n(x)=\sqrt{n(x)}$, we get a final formula for the errors:
\begin{equation}
\delta a_\alpha^2 = \sum_{\textrm{event }n} \sum_{\beta,\gamma} M_{\alpha\beta}^{-1}\frac{d\ln P(x_n)}{da_\beta}M_{\alpha\gamma}^{-1}\frac{d\ln P(x_n)}{da_\gamma} = N\left(M^{-1}CM^{-1}\right)_{\alpha\alpha}
\end{equation}
Where $C$ is the covariance matrix: $C_{\alpha\beta} \equiv \left<\partial_{a_\alpha} \ln P \partial_{a_\beta} \ln P \right>$.

In the simple case of only a single $a$, this reduces to:
\begin{equation}
\delta \hat{a}= 
\frac{\sqrt{\sum_n(\frac{\partial}{\partial a}\ln P(x_n;\hat{a}))^2}}{\frac{\partial^2\ln\mathcal{L}(\hat{a})}{\partial a^2}}
\end{equation}

This formalism can be extended to understand the use of the EML for multiple event interpretations. 
For multiple event interpretations, correlations between points can be resolved by
expressing the sum over $K$ events as a sum over the multidimensional parameter space 
of $\vec{x}=\{x^1, x^2,..., x^I\}$, where the new index refers to each event interpretation. For the analysis
where all interpretations are combined in the same $x$ distribution, $P(x)=\sum_i P_i(x)$ and
$P_i(x)$ is the projection of the multidimensional distribution $P(\vec{x})$ along the 
interpretation $i$. 

In this approach the likelihood function to be maximized is:
\begin{equation}
\ln{\mathcal{L}}=\sum_{k=1}^K \sum_{i=1}^I\ln P(x^i_k;\{a_\alpha\})-I\mathcal{N}(\{a_\alpha\}). 
\end{equation}
Computing the errors from this point of view is complicated because the $x^i_m$ are correlated. But we can recast it like this:
\begin{equation}\label{multidimlikelihood}
\ln{\mathcal{L}}=I \left(\sum_{k=1}^K \ln \tilde P(\vec{x}_k;\{a_\alpha\})-\mathcal{N}(\{a_\alpha\})\right)
\end{equation}
Were $\ln\tilde P(\vec{x}) \equiv \frac{1}{I}\sum_i \ln P(x^i)$. Now Eq. \eqref{multidimlikelihood} looks exactly like Eq. \eqref{onedimlikelihood}, and the derivation of the errors is exactly the same, except using $\tilde P$. Thus we have:
\begin{align}
\delta a_\alpha^2 &= K \left(M^{-1}CM^{-1}\right)_{\alpha\alpha}
\\
M_{\alpha\beta} &= \frac{d\ln\mathcal{L}/I}{da_\alpha da_\beta}
\\
C_{\alpha\beta} &= \frac{1}{I^2}\left<\frac{d\sum_i \ln P(x^i)}{da_\alpha}  \frac{d \sum_i\ln P(x^i) }{da_\beta} \right>
\end{align}
And in the case of a single fit parameter $a$:
\begin{equation}
\delta \hat{a}= 
\frac{\sqrt{\sum_k\left(\frac{\partial}{\partial a}\sum_i\ln P(x^i_k;\hat{a})\right)^2}}{\frac{\partial^2\ln\mathcal{L}(\hat{a})}{\partial a^2}}
\end{equation}
In the case of interest, we have a signal distribution $\rho_S(x)$ and a background distribution $\rho_B(x)$ (each normalized so $\int \rho(x) dx$ is the number of interpretations expected to pass the cuts). We are fitting to a predicted distribution $P = N_S\rho_S+N_B\rho_B$, so we have two parameters $N_S$ and $N_B$. The likelihood is:
\begin{equation}
\ln{\mathcal{L}}=\sum_{k=1}^K \sum_{i=1}^I \ln (N_S\rho_S(x_k^i)+N_B\rho_B(x_k^i))-R(N_S+N_B). \label{logL}
\end{equation}
And then the errors are given by:
\begin{align}
\delta N_\alpha^2 &= N\left(M^{-1}CM^{-1}\right)_{\alpha\alpha}\label{errresult}
\\
M_{\alpha\beta} &= \sum_{k=1}^K\frac{1}{I}\sum_{i=1}^I \frac{-\rho_\alpha(x_k^i)\rho_\beta(x_k^i)}{(N_S\rho_S(x_k^i)+N_B\rho_B(x_k^i))^2}
\\
C_{\alpha\beta} &= \left<\left(\frac{1}{I}\sum_{i=1}^I\frac{\rho_\alpha(x_k^i)}{N_S\rho_S(x_k^i)+N_B\rho_B(x_k^i)}\right)\left(\frac{1}{I}\sum_{i=1}^I \frac{\rho_\beta(x_k^i)}{N_S\rho_S(x_k^i)+N_B\rho_B(x_k^i)}\right)\right>\label{errresultB}
\end{align}
Using Eqs.~\eqref{errresult}-\eqref{errresultB} and \eqref{sigfit}, the improvement in the significance of the search can be estimated for
an analysis using a likelihood fit. 

\begin{figure}[t]
  \begin{center}
  \subfigure[][]{
\includegraphics[width=0.48\textwidth]{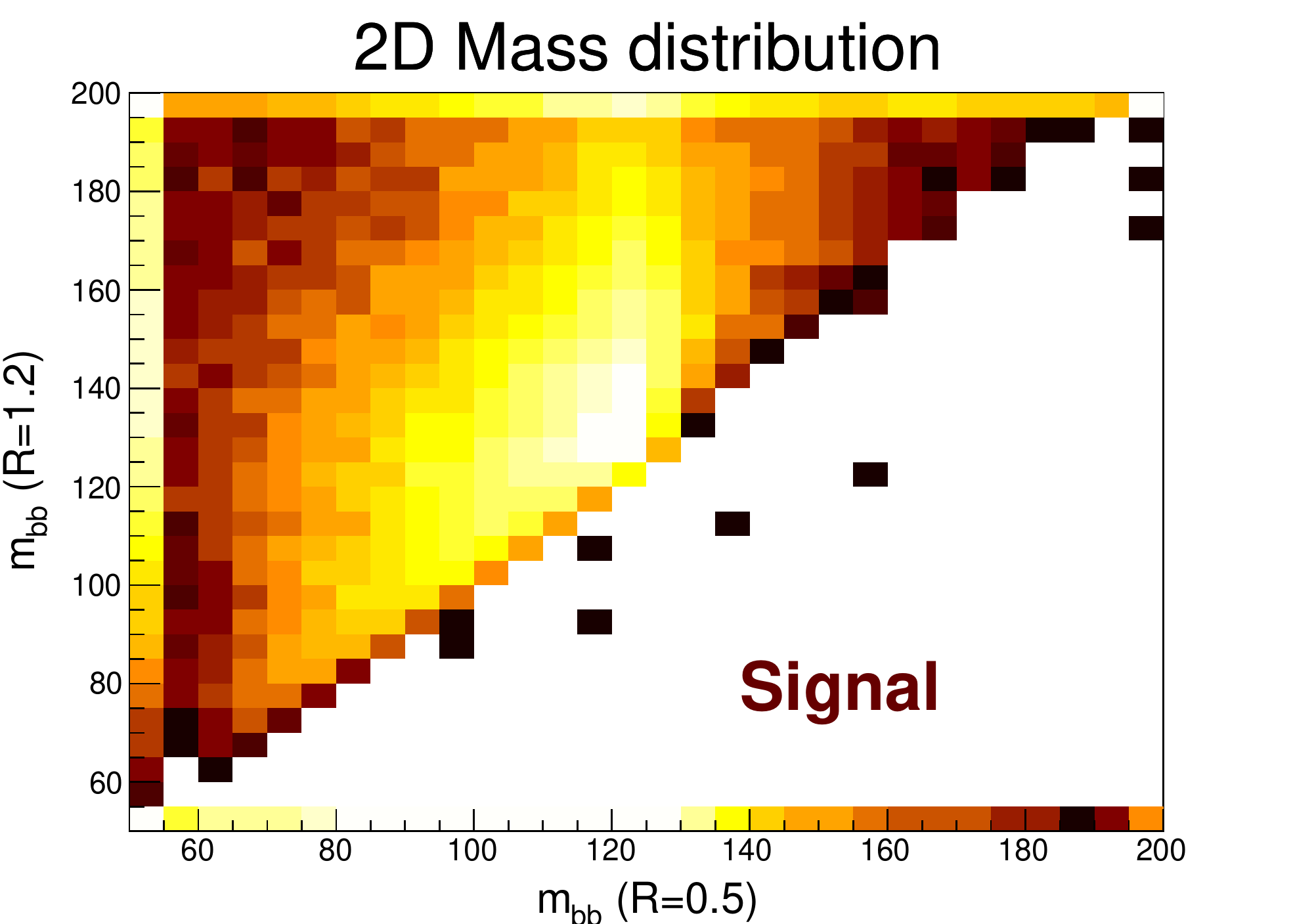}
}
\hspace{-5mm}
  \subfigure[][]{
\includegraphics[width=0.48\textwidth]{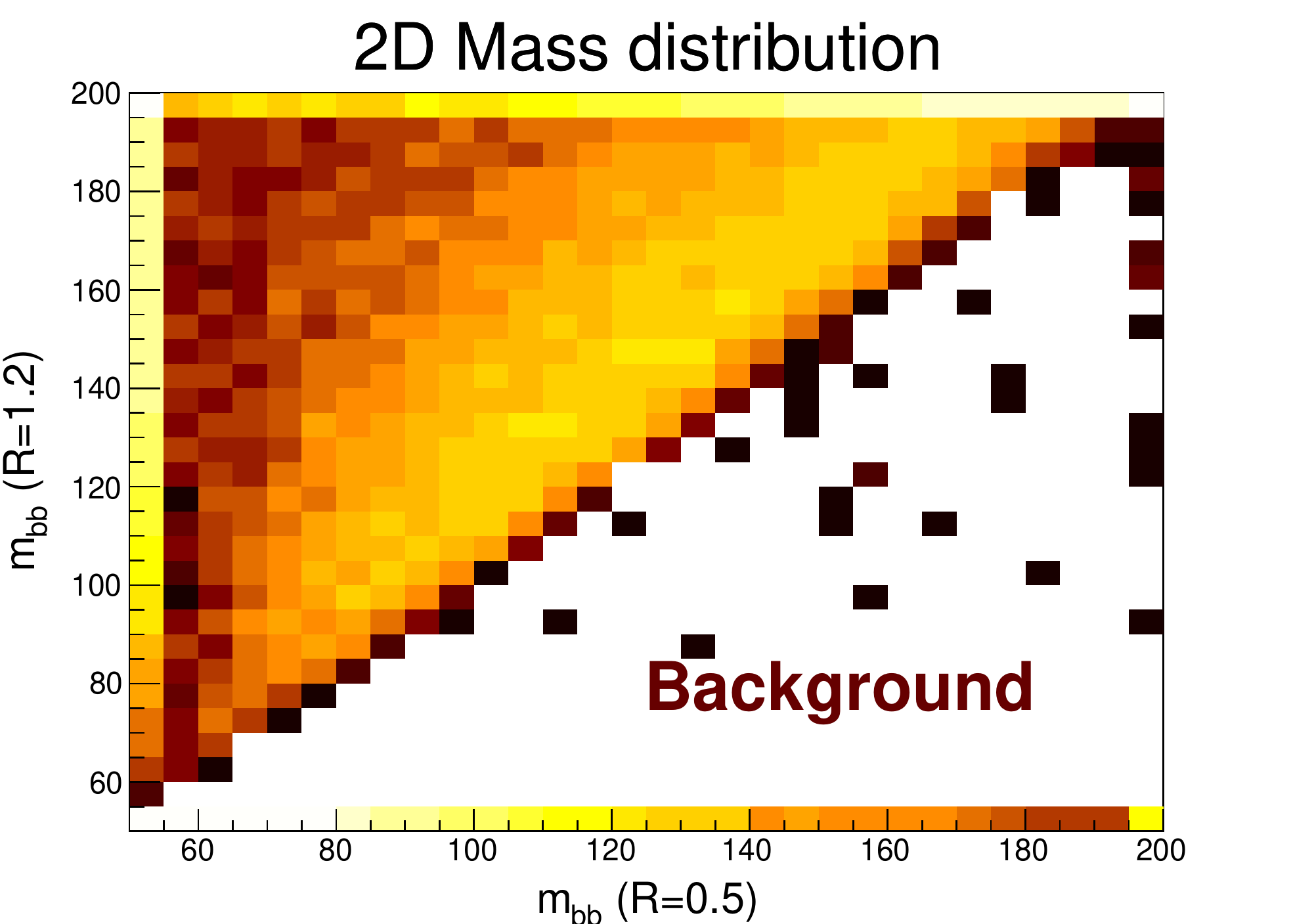}
}
\caption{Two-dimensional distributions of invariant masses reconstructed with the anti-$k_t$ 
algorithm with $R=0.5$ and $R=1.2$ for events with $p_T^Z>120$~GeV for the signal (a)
and background (b). Colors indicate relative cross sections, with lighter colors being
more probable. Normalization is arbitrary.
}
\label{fig:2Ddist}
  \end{center}
\end{figure}

\subsection{Multidimensional likelihood fits}\label{sec:multLike}
Rather than merging all the interpretations into a single distribution, it would clearly be smarter to keep the interpretations
separate and exploit their correlations. Unfortunately, computing the exact likelihood from an $I$-dimensional space would require
an exponentially large data set. For example, with $I=10$ interpretations, and only $B=5$ bins in each direction, we would need 10 million events just to have
each bin populated with 1 event. This is of course just the usual curse of dimensionality for multivariate fits, which is present in any analysis with correlated
observables. A popular approach is to replace the exact likelihood with boosted decision trees (BDTs), neural networks, or other sophisticated algorithms. 
Another approach is just to take $I$ and $B$ small enough so that the dimensionality is not intractable. In this section, we compare these alternatives.

\begin{figure}[t]
  \begin{center}
   \subfigure[][]{
\includegraphics[width=0.3\textwidth, angle=90]{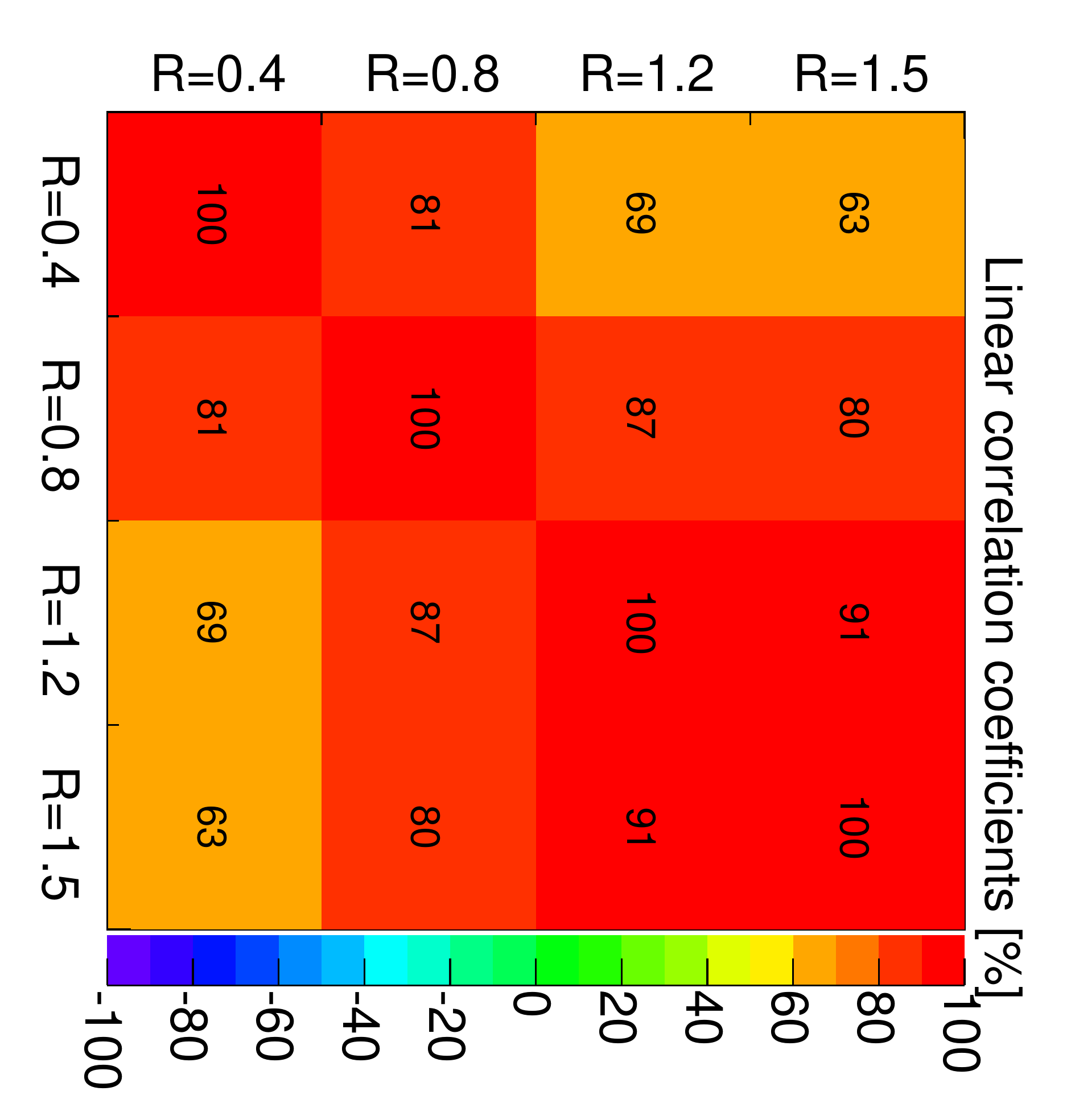}
}
  \subfigure[][]{
\includegraphics[width=0.3\textwidth, angle=90]{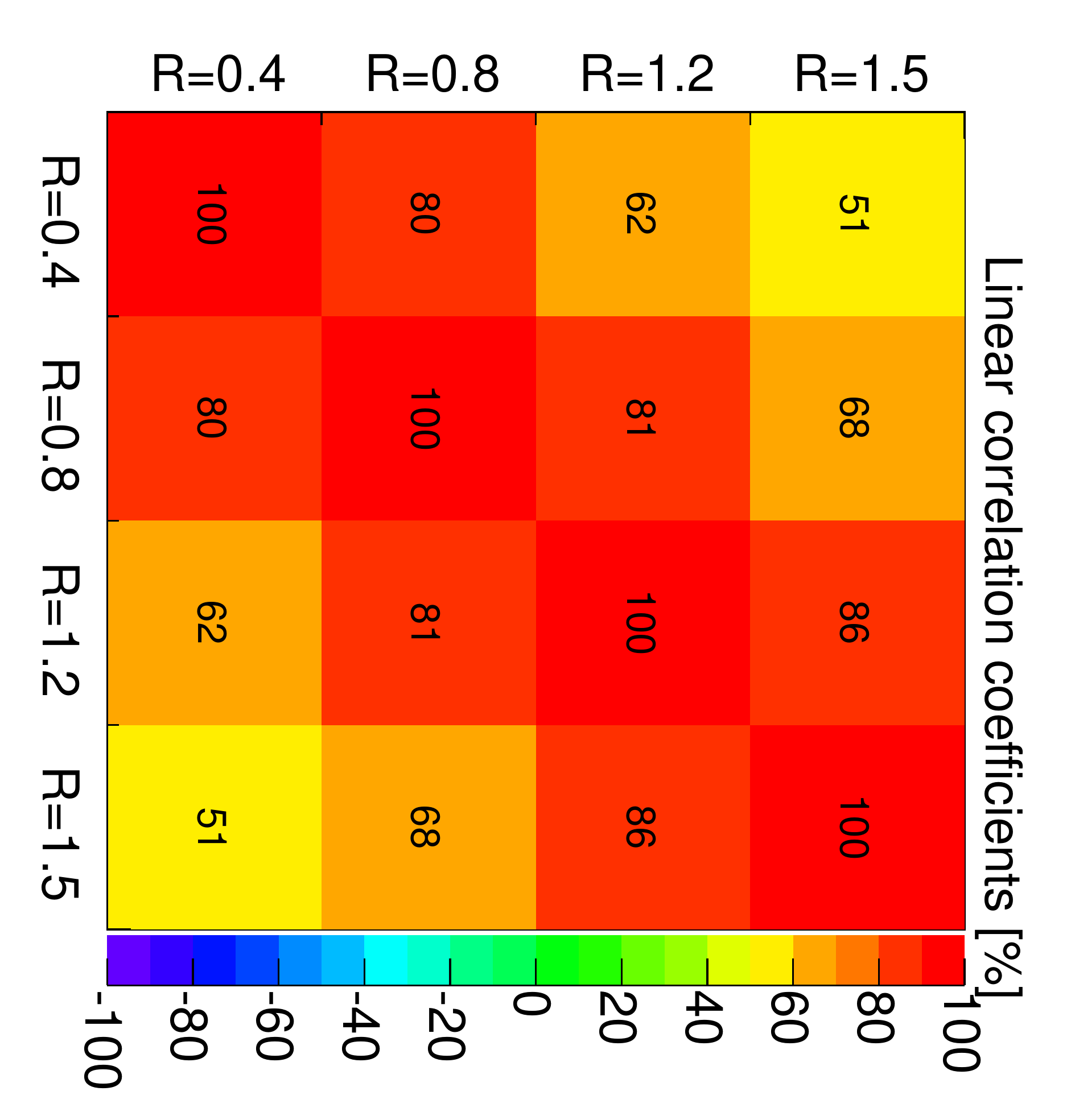}
}
  \subfigure[][]{
\includegraphics[width=0.3\textwidth, angle=90]{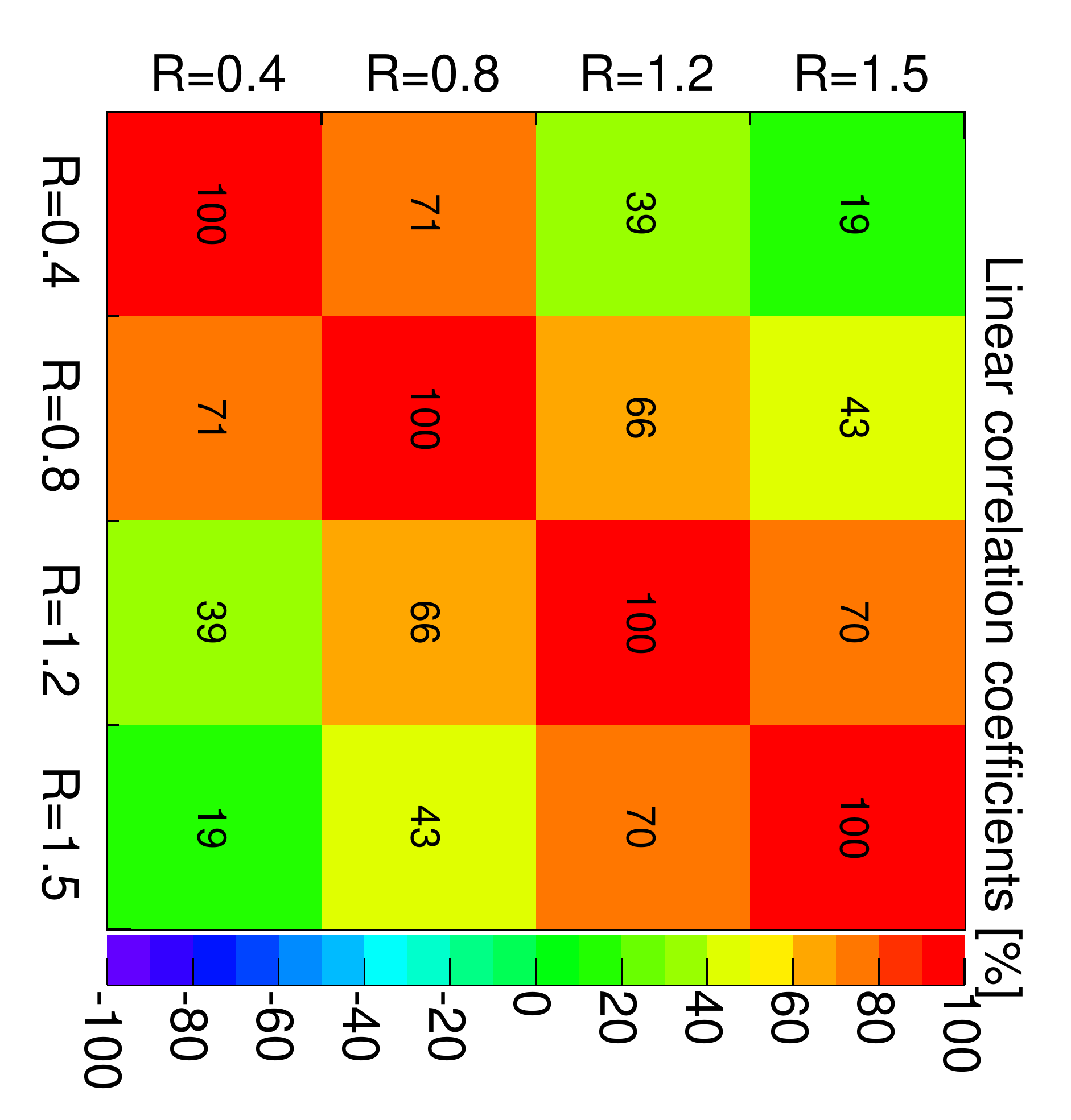}
}
  \subfigure[][]{
\includegraphics[width=0.3\textwidth, angle=90]{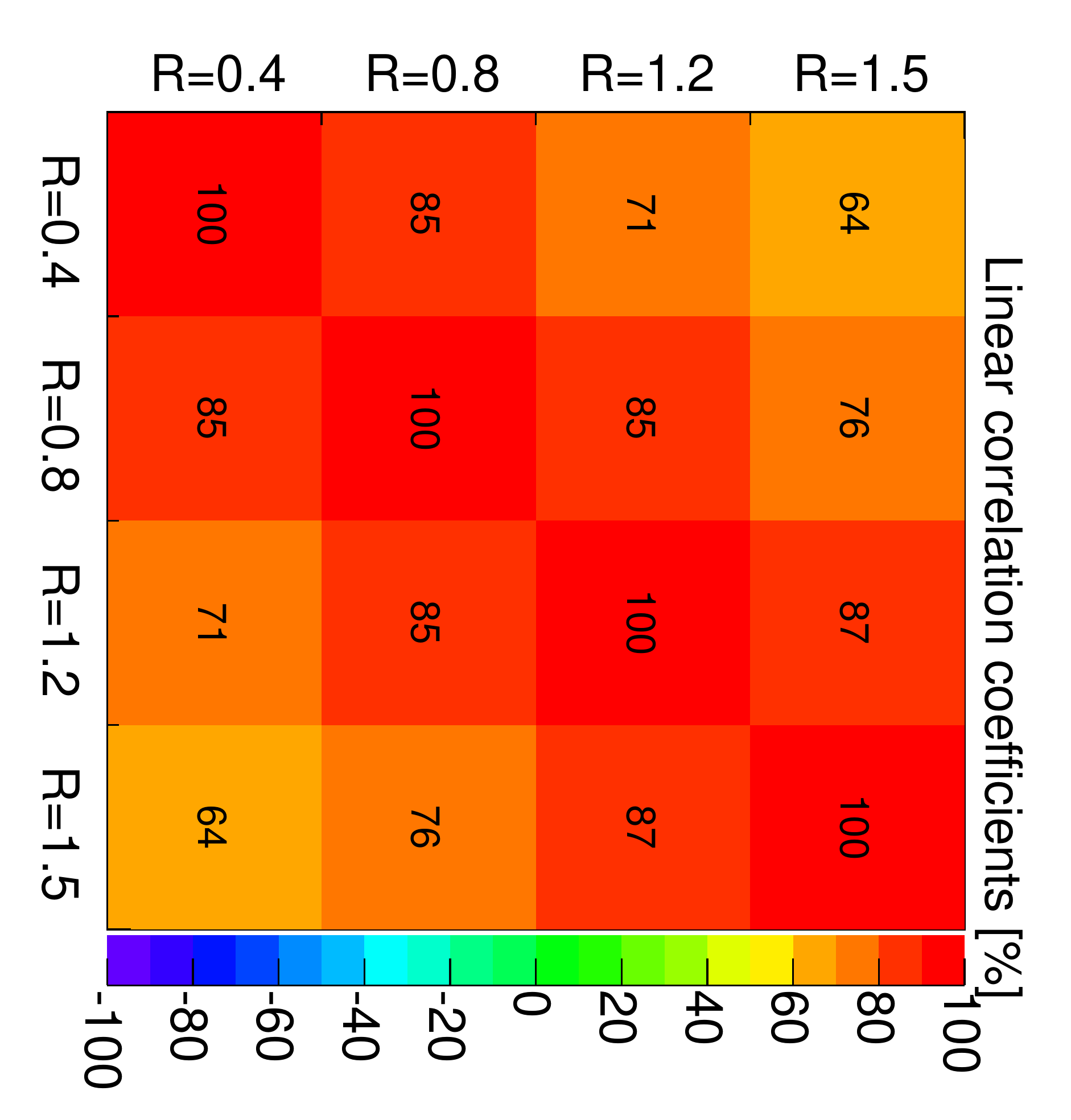}
} 

\caption{Linear correlation coefficients for $\mbb$ among different $R$ values for
signal (a,c) and background (b,d), with $p_T^Z<120$~GeV (a,b) and  $p_T^Z>120$~GeV on bottom (c,d).}
\label{fig:correlations}
  \end{center}
\end{figure}

If we take $I=2$, then it is possible to populate the 2 dimensional space quite well. For example, 
 the 2D distributions for signal and background $\mbb$ distributions for two different choices of $R$ are shown in Figure~\ref{fig:2Ddist}.
The figure shows the low $p_T^Z$ sample clustered with anti-$k_t$ using $R=0.5$ and $R=1.2$. 
The results at high $p_T^Z$ are qualitatively comparable except for an overall shift at 
higher masses for the background. 
The two invariant masses are clearly correlated, but also clearly not 100\% correlated.
To quantify the correlation, the linear correlation coefficients among  some representative $R$ values are shown in Figure~\ref{fig:correlations}.
A picture emerges in which, particularly at high $p_T^Z$, the correlation between
small $R$ and large $R$ interpretations is quite different for signal and background,
with smaller correlations in the signal. The 2-dimensional data can be run through a regular likelihood fitting procedure, with no modification since each event is only contributing a single point.
We show results in Table~\ref{tab:results}.

\enlargethispage{10pt}
As an alternative to taking $I=2$, we can produce a statistically tractable fitting problem by instead reducing the number of bins $B$. For instance, 
we consider $B=3$ and $I=4$, where we bin every interpretation into $\mbb < 110$ GeV, $110$ GeV $< \mbb < 140$ GeV, and $140$ GeV $< \mbb$. Results
for this approach are also shown in Table~\ref{tab:results}.

\begin{table}[htb]
\centering
\begin{tabular}{|c|c|c|c|c|}
      \hline
      Improvement  & \multicolumn{2}{c|}{$p_T^Z<120$~GeV}  & \multicolumn{2}{c|}{$p_T^Z>120$~GeV}  \\ 
      (over R=0.5 only) & xs-based & EML & xs-based & EML  \\
      \hline
	1 $R$,   fraction in window & 0.83 & 0.74 & 0.80 & 0.71\\
      12 $R$'s, fraction in window & 0.98 & 0.97 & 0.92 & 0.88 \\
	1 $R$, $\mbb$& 1.00 & 1.00 & 1.00 & 1.00 \\
      5 $R$'s, $\mbb$ merged & 0.94 & 1.08 & 0.94 & 1.06 \\
      4 $R$'s, 3 bins & 0.99 & 1.00 & 1.16 & 1.20 \\
      2 $R$'s, full & 1.10 & 1.14 & 1.35 & 1.38 \\
       2 $R$'s, BDT & 1.04 & 1.08 & 1.30 &  1.34 \\
       12 $R$'s, BDT& 1.19 & 1.30 & 1.52 & 1.41 \\
\hline
      12 kinematic & 1.33 & 1.50 & 1.35 & 1.29 \\
      12 kinematic + 12 $R$'s & 1.39 & 1.68 & 1.67 & 1.55 \\
\hline
\end{tabular}
    	\caption{
\label{tab:results}
Relative significances are shown for various collections of jet sizes and the different methods discussed in the text.
This table is explained in Section~\ref{sec:results}.
} 
\end{table}

\subsection{Results \label{sec:results}}

Table~\ref{tab:results} summarizes the results obtained with the different methods described 
in previous sections. 
The baseline is the third line of the table; for the xs-based results it uses $\rho(\mbb)$ (see Section~\ref{sec:measures}) and for the EML-based result uses a fit to the $\mbb$ distribution.
The first two lines use the fraction of interpretations in a window of $110$~GeV~$< \mbb <$~140~GeV to define $z$ as in 
~\cite{Chien:2013kca}. These two rows list the xs-based significance and the EML significance, both evaluated on the observable $z$ given by the fraction of interpretations that lie in the window. For 1 $R$, $z$ is either 0 or 1, and this is equivalent to a simple cut-and-count experiment. 
The use of the full $m_{bb}$ distribution (from 50 GeV to 200 GeV with an overflow bin on each side), even
with just 1 $R$ (the third row) is more powerful than the use of multiple event interpretations for the 
simple cut-and-count model (the second row). When using 5 $R$'s ($R=0.4,0.5,0.6,0.7$ and $R=0.8$) and pooling
all interpretations in the same $\mbb$ distribution (the row labeled ``merged''), there is a small gain when using the likelihood fit, despite 
some features of the distribution being washed out in the procedure. 

The next row shows the results when using 3 $\mbb$ bins ($\mbb<110$~GeV, $110$~GeV~$< \mbb <$~140~GeV and 
$\mbb>140$~GeV),  to reduce the dimensionality, and 4 $R$'s ($R=0.4, 0.6, 0.8, 1.0$). Using 3 bins provides non-negligible
gains at high $p_T$, but does not
manage to perform better than a single interpretation at low $p_T$. This indicates that at low $p_T$ the added bins do not help tell
apart the different interpretations but that a $p_T$-dependent choice of binning might be worth further exploration.

The method that gives the second highest significance is the one using the full 2D distributions of $\mbb$ computed with $R=0.5$ and $R=1.2$. At high $p_T$, we find  35\% and 38\% improvements with the xs-based and EML
fit respectively. As long as the two radii are far enough apart that 
the correlations between the reconstructed invariant masses is not very high, the choice of the 
two radii does not impact significantly the observed improvement. The use of boosted decision trees\footnote{We use the default BDT parameters in the TMVA package. The significance  is computed using the probability-distribution-functions produced from TMVA from the BDT classifier~\cite{Therhaag:2009dp}.} to combine 
the two radii does not do quite as well as the full 2D $m_{bb}$ distribution, suggesting that there is a loss of information in the construction of the BDT, at least in our implementation (the TMVA default). 
\enlargethispage{10pt}

The improvement in both the xs-based and the EML-based approaches is highest using the most radii which we combine using BDTs. We find up to a 52\% improvement  for xs-based or 41\% for EML based in the high $p_T$ sample by combining 12 $R$'s over using just a single $R$. Two additional entries in the table refer to the use of kinematic variables in the BDTs and are discussed
in Section~\ref{sec:mva}. 

\section{Comparison to standard observables}\label{sec:mva}

So far, we have seen that combining measurements of $\mbb$ computed by clustering with the anti-$k_t$ algorithm with different $R$'s can have a sizable improvement over using $\mbb$ with a single $R$.
It is natural to wonder whether the improvement is due
to the exploitation of properties of the events which could be exploited equally well using more traditional kinematic variables. 
For example, as $R$ increases, the jet momentum increases. Since the signal and background have different momentum dependence, one might imagine
that the same gain could be realized simply by including the $p_T$ of the $b$-$\bar{b}$ system into the analysis.
To explore whether multiple $R$ values leads to improvement in the $H Z \to b\bar{b} \ell^+ \ell^-$ search, we compare the improvement using
multiple event interpretations (multiple $R$'s) to the improvement from standard kinematic observables.

For the kinematic variables, we take the set from the $HZ$ LHC study in~\cite{Gallicchio:2010dq}, Table 4. 
All observables
are computed using jets reconstructed with the anti-$k_t$ algorithm with $R=0.7$. The observables
we consider are constructed from either the hadrons ($b$-jets):
\begin{equation}
\mbb,
\quad p_T^{b_1},
\quad p_T^{b_2},
\quad |p_T^{b_1}|+|p_T^{b_2}|, 
\quad |\eta^{b_1}-\eta^{b_2}|, 
\quad |\eta^{h}-\eta^{b_2}|,
\quad\text{and}\quad
|p_T^{h}-p_T^{b_2}|
\end{equation}
or the leptons (from the $Z$ decay):
\begin{equation}
p_T^{\ell_1},
\quad
p_T^{\ell_2},
\quad 
p_T^{Z}
\quad\text{and}\quad
m_{\ell\ell}
\end{equation}
and one variable dependent on the leptons and $b$-jets:
\begin{equation}
|\eta^h - \eta^Z|
\end{equation}
In these expressions, $\eta$ is rapidity, $p_T$ is transverse momentum, $b_1$ refers to the harder of the two $b$-jets and $b_2$ to the softer, and similarly for the leptons.

These 12 variables are all input to a BDT and used to compute the significance just as we have for multiple event interpretations.
We also try combining these kinematic variables with the 11 remaining masses in our study (computed using $R=0.4$ to $R=1.5$ in steps of $0.1$).
Results are shown on the last two lines of Table~\ref{tab:results}. 
We can draw a few interesting qualitative conclusions from this analysis. First, we see that the kinematic variables work 
relatively better at low $p_T$ than high $p_T$. At high $p_T$ angular differences are smaller and there are, thus, fewer 
handles to distinguish signal from background. This is unlike the multiple event interpretations, for which improvements are
more significant at high $p_T$. Second, we see that using multiple event interpretations (multiple $R$'s) 
still gives serious benefit on top of all of the kinematic variables. The improvement is more significant at high $p_{T}$ as
expected from the previous observation. 

\section{Conclusions}\label{sec:concl}
In previous work~\cite{Ellis:2012sn,Kahawala:2013sba,Chien:2013kca} multiple event interpretations, in particular the reclustering of an event using different jet sizes, were shown to
give sizable improvement in the potential significance for a $H\to b \bar{b}$ search in association with a $W$ or $Z$ boson. 
In this paper we have attempted to refine those analyses using methodologies as close as possible to those used in  experimental 
analyses at the LHC. For this purpose, a new expression of the likelihood
has been developed to account for correlations across events populating
several bins in one dimension. 
The improvement in the significance of the search 
with this treatment has been shown to be sizable. We find as much as $\approx 41\%$ improvement when using 12 $R$'s over a single $R$ when the $b\bar{b}$ system has $p_T > 120$ GeV and $\approx 30\%$ improvement for $p_T < 120$ GeV. 
We also explored whether the improvement from multiple event interpretations carries overlapping information  to traditional kinematic variables or complementary information. To answer this question, we took 12 kinematic variables that have been
demonstrated to be nearly optimal in a multivariate analysis~\cite{Gallicchio:2010dq} (and some of which were used in a recent CMS search) for $H\to b\bar{b}$ and compared their efficacy to what we get from just using multiple $R$'s and what we get by combining them. We find that adding the $m_{bb}$ at multiple interpretations gives a $12\%$ improvement
at low $p_T$ and $20\%$ improvement at high $p_T$ over the kinematic variables at a single $R$.
These improvements are particularly encouraging, since the phase space explored in this paper
does not include boosted topologies and thus cannot benefit from otherwise highly successful 
jet substructure techniques~\cite{Butterworth:2008iy, Kaplan:2008ie, Cui:2011xy}. This phase space is, however, quite relevant for finding the $H \to b\bar{b}$ decay.
In summary, these results strongly suggest that multiple interpretations can help in searches with realistic statistical methods.

\acknowledgments
MDS and DF are supported in part by the DOE under grant DE-SC003916. YTC is supported by the US Department of Energy, Office of Science. 
AM is supported in part by a DoD NDSEG fellowship.

\bibliography{bib}
\bibliographystyle{jhep}
\end{document}